# Tipping Elements in the Human Intestinal Ecosystem

Leo Lahti[1,2*], Jarkko Salojärvi[1,†] Anne Salonen[3,†], Marten Scheffer[4], Willem M. de Vos[1,2,3]

[1]Department of Veterinary Biosciences, University of Helsinki, P.O. Box 66, FI-00014, Helsinki, Finland. [2]Laboratory of Microbiology, Wageningen University, P.O. Box 8033, 6700 EJ Wageningen, The Netherlands. [3]Department of Bacteriology and Immunology, Haartman Institute, University of Helsinki, P.O. Box 21, FI-00014 Helsinki, Finland. [4]Aquatic Ecology, Wageningen University, P.O. Box 47, 6700 AA Wageningen, The Netherlands. [†]These authors contributed equally to this work *Corresponding author:  Leo Lahti (leo.lahti@iki.fi)

## Abstract

Recent studies show that the microbial communities inhabiting the human intestine can have profound impact on our well-being and health. However, we have limited understanding of the mechanisms that control this complex ecosystem. Based on a deep phylogenetic analysis of the intestinal microbiota in a thousand western adults we identified groups of bacteria that tend to be either nearly absent, or abundant in most individuals. The abundances of these bimodally distributed bacteria vary independently, and their contrasting alternative states are associated with host factors such as ageing and overweight. We propose that such bimodal groups represent independent tipping elements of the intestinal microbiota. These reflect the overall state of the intestinal ecosystem whose critical transitions can have profound health implications and diagnostic potential.

Recent studies show that the diverse microbial communities inhabiting the human intestine can have profound impact on our well-being and health[1-4]. Although compositional and functional properties of the intestinal microbiota have been studied extensively, we have limited understanding of the mechanisms that control this complex ecosystem[5-6]. A fundamental question is whether the overwhelming variability of the microbiota composition occurs along phylogenetic gradients, or through abrupt switches between alternative stable states[7]. In ecology, the concept of alternative stable states separated by tipping points has become a cornerstone of the theoretical framework with major practical implications[8]. Our study now demonstrates that the intestinal ecosystem has the same properties. While previous studies have focused on analyzing ecosystem-wide configurations in human intestinal microbiota[7,9-10], our analysis of intestinal microbiota in a thousand western adults shows that there are distinct groups of bacteria that tend to be either very abundant or very rare. This bimodality indicates that these groups have alternative stable states. We show that host factors such as age can affect the resilience of the alternative states and move the system towards a tipping point. We propose that the bimodal groups we identified represent 'tipping elements'[11] linked to our physiology, and that resetting these 'bacterial dip switches' may be a radically new way to approach the rapidly growing number of health issues related to the intestinal microbiota. This will likely change the way we look at management of the intestinal ecosystem.

**Results**

To systematically investigate this hypothesis, we scanned 130 genus-like phylogenetic groups that cover the majority of the known bacterial diversity of the human intestine[12] (Supplementary Table 1) to assess evidence for alternative states in these bacteria. We used the potential analysis methodology that was recently introduced to study analogous state shifts in climatology and forest ecosystems[13-14], as it provides a robust approach for the identification of alternative states in noisy dynamical systems, such as the remarkably variable intestinal ecosystem. Our analysis is based on >2000 standardized phylogenetic microarray hybridizations that represent the fecal microbiota of thousand adults (n=1006; 18-77 years) from 15 western countries (Europe and US) devoid of reported health complications[15]. Our analysis complements and extends the earlier studies by providing increased phylogenetic depth[15] and sample size.

Based on this extensive data collection, we identified several bacterial groups exhibiting robust bimodal abundance distributions (Fig. 1; Supplementary Table 2); our analyses confirmed consistent bimodality in *Dialister spp.*, relatives of *Bacteroides fragilis (B.fragilis), Prevotella melaninogenica (P.melaninogenica)*, *Prevotella oralis (P.oralis),* and two groups of the Uncultured Clostridiales (UCI and UCII)[11]. The bimodality of these groups was robust against the choice of the DNA extraction method and bootstrap sampling of the data, and was supported by different methodologies[14]. The other 124 genus-like groups that we investigated were either rare or exhibited symmetric or skewed abundance distributions (Supplementary Fig. 1; Supplementary Table 1). Although the bimodally distributed bacteria tend to be either rare or abundant in most subjects, intermediate states were also frequently observed (Fig. 1). These may correspond to subjects in a transitional phase between the states for instance due to changes in diet or physiological state (Supplementary Fig. 2).

Remarkably, most bimodal groups exhibited only weak preference for mutual co-occurrence or exclusion (Pearson $|r| < 0.12$; Fig. 2). Exceptions included a positive correlation between UCI and UCII (r=0.50), and between *P.melaninogenica* and *P.oralis* (r=0.97). We combined the highly correlated *Prevotella*s into a single group. The low- and high-abundance states of the resulting five bimodal groups yield altogether 32 ($2^5$) possible combinations with varying population frequencies. The bimodal groups co-varied with other bacteria (Supplementary Fig. 3), suggesting that state shifts in the bimodal groups indicate wider changes in the overall ecosystem composition. While *Prevotella* and *Bacteroides* have been earlier identified as drivers of mutually exclusive community-level enterotypes[9], our observations pinpoint specific bimodal bacteria within these groups and highlight further, previously unreported bimodal groups. The independent variation and frequent co-occurrence of these groups indicates that their alternative states are complementary rather than mutually exclusive.

The emerging picture is that rather than overarching alternative states corresponding to distinct community types, the natural variation of the human intestinal microbiota can be associated with specific bacterial groups representing *'tipping elements'* that are only loosely coupled, analogously to the climate system that is now also believed to have distinct tipping elements[11]. An important corollary is that the resilience (the capacity to recover from perturbations) of the alternative states will be influenced by a range of factors that may selectively affect specific bacterial groups. Hence, we should expect that gradual changes in external or internal factors may drive the implicated bacteria towards a tipping point where an abrupt shift between the contrasting alternative states may follow[8,16]. One way to infer how resilience of the microbiota might depend on such factors is to use probability densities of bacterial abundance to estimate the approximate basins of attraction. Based on this we can observe, for instance, that a basin of attraction around an apparently resilient high-abundance state of UCI arises in late middle age (Fig. 3; Supplementary Table 3). This suggests that with increasing age, the resilience of the low-abundance state decreases and eventually even a small perturbation may induce a shift to the high abundance state. Moreover, the analysis shows that population-level studies may average out alternative states that are pronounced in specific subpopulations, such as particular age groups (Fig. 3B).

Overall, we observed a range of associations between the state probabilities and host factors such as ageing, overweight and health status[15]. Analysis of an additional phylogenetic microarray data set derived from subjects with irritable bowel syndrome and severe obesity indicated association between these health complications and the low-abundance UCI and UCII. Metabolic syndrome was positively associated with *B.fragilis* (false discovery rate FDR<5%), and showed trends (FDR<20%) to a positive association with *Dialister,* and a negative one with *Prevotella* (Supplementary Table 4). Furthermore, we observed an association between the bimodal bacteria and the overall gene richness of the microbial community (Wilcoxon test p<0.01 for all groups except *B.fragilis*; Supplementary Fig. 4) when we incorporated the recently reported metagenomic sequencing data for 255 subjects[17]. In our analysis of these meta-genomes, the high-abundance *B.fragilis* group and *Dialister spp.* were found to be associated with a low gene count, in line with their observed association with pronounced metabolic dysfunction[17]. On the contrary, we observed associations between *Prevotella*, UCI and UCII and the high gene count, which has been linked to a potentially healthy metabolic phenotype[17]. These results demonstrate that differences in health status are often reflected in changing state probabilities of specific taxonomic groups.

To complement our cross-sectional analysis that represents a static snapshot of the intestinal ecosystem in each individual at a specific point of time, we analyzed short-term temporal stability of the observed states in 78 individuals (27-63 years) across a three-month interval. The more frequent states were in general more stable (Supplementary Fig. 5), as expected based on stationary-state continuous Markov process (Supplementary Fig. 6) and consistent with the idea that changes in species abundance distributions can be used to infer changes in the stability and resilience of alternative states[18]. The high-abundance *Prevotella* was a notable exception, however, as one of the least frequent yet one of the most stable states. Further intervention studies with experimentally induced state shifts will be needed to address the causal relations and long-term stability, and to investigate whether these states translate into differential disease susceptibility or drug response of the host.

**Discussion**

Bimodality in bacterial abundances can be induced by many factors, such as disruptive selection[19], competition and cooperative feedback loops[20], or environmentally or genetically determined host factors. Our results demonstrate that despite the dominating nature of gradual variation in bacterial abundances in the intestinal ecosystem, specific bacterial groups form a limited number of contrasting, stable configurations that associate to host physiology and health. The bimodal taxa, together with other microbial co-occurrence relationships[21-23], will be instrumental in providing insight on the variation, regulation and health implications of the intestinal microbiota. The implicated taxa may allow stratification of the subjects based on alternative states that in parallel to gene richness[17] represent promising novel targets for microbiome-based diagnostics and therapies[24]. Targeting specific subpopulations, opposed to the daunting complexity and variability of the entire ecosystem, can simplify the characterization and possible manipulation of the intestinal microbiota.

**Methods**

**Sample collection** Over 5000 intestinal samples were available in the in-house MySQL database of the Human Intestinal Tract (HIT) Chip, a phylogenetic microarray[12]. To investigate the microbiota variability in western adult population, we selected from the HITChip database 1006 western adults (18-77 years; average body-mass index 26.7 kg/m$^2$; standard deviation 5.8 kg/m$^2$) with no reported health problems. A single fecal sample per subject was selected; the first time point was used if multiple samples were available. The subjects cover 15 western countries (Europe and US) that were further aggregated into Central European (Belgium, Denmark, Germany, Netherlands),  Eastern European (Poland), Nordic (Finland, Norway, Sweden), South European (France, Italy, Serbia, Spain), UK/Ireland (UK, Ireland), and US (US) groups.

**Microbiota profiling with phylogenetic microarrays** Culture-independent techniques for characterizing intestinal microbiota include phylogenetic microarrays and sequencing-based approaches[25]. Phylogenetic microarrays can be used to quantify with high accuracy the known phylotypes including the low-abundance phylotypes that often remain undetectable with conventional sequencing depths[26]. Our analysis is based on the phylogenetic HITChip microarray[11], produced by Agilent Technologies, Inc. (Santa Clara, California, USA), that provides a sensitive analysis platform to assess differences in relative abundance of intestinal bacteria based on fecal samples. The HITChip microarray targets the V1 and V6 hypervariable regions of 16S rRNA gene, and can detect 1033 species-like bacterial phylotypes (> 98% sequence similarity in the 16S rRNA gene) that re-present the majority of the bacterial diversity of the human intestine[12]. For the analyses, hybridization signals were summarized to 130 genus-like phylogenetic groups (>90% sequence similarity in the 16S rRNA gene) that are referred to as *Species* and relatives, the latter being shortened as "*et rel.*"[12]. The HITChip microarray allows highly reproducible (98±2% Pearson correlation across technical replicates) and deep analysis (reproducible detection of phylotype abundances below a 0.1% relative abundance level) of the phylotype composition of intestinal samples, comparable to 200,000 next generation sequencing reads per sample[26]. Hence, the standardized analytical pipeline for HITChip microarray data collection provides advantage over sequencing-based approaches, where the comparable collections of intestinal microbiota profiling data are smaller, and integration of data from individual studies is more challenging e.g. due to use of different primers, continuously changing sequencing platforms and protocols. The fecal samples, collected in the context of clinical trials, were typically taken at home, frozen at -20ºC, delivered to the study center within 24 hours and stored at -80ºC. The DNA was extracted and prepared by amplification of the full-length 16S rRNA gene, followed by transcription and labeling of the resultant RNA with Cy3 and Cy5 prior to fragmentation and hybridization on the array[12]. The signal intensity data from the microarray hybridizations was collected using the G2505C scanner by Agilent and preprocessed using an in-house MySQL database and custom R scripts. Each scanner channel from the array was spatially normalized using polynomial regression, followed by outlier detection and filtering in each set of probes with a $\chi^2$ test. The intensities were sample-wise quantile normalized and averaged to a single value for each probe-sample pair. Each sample was hybridized on at least two replicates to ensure reproducibility. Between sample normalization was performed at probe level with the minimum-maximum normalization using 0.5% quantiles estimated from the data. Each targeted genus-like group is probed by multiple (median 19) oligos. Logarithmized ($\log_{10}$) signal of the probes targeting the same phylotype were summarized with fRPA[27] (*RPA* R package). These log-transformed signals were used as a proxy for bacterial abundance.

**Bacterial abundance types** We categorized the 130 genus-like groups into characteristic '*abundance types*' to gain an overview of their overall population-level variability (Supplementary Fig. 1; Supplementary Table 1). 60 groups were abundant and prevalent, exceeding HITChip $\log_{10}$ signal of 3.0 in >20% of the subjects; the remaining 70 groups were considered rare. The observed abundance types included symmetric, skewed and bimodal distributions. The 34 prevalent groups with symmetric log-abundance (absolute skewness <0.5; *e1071* R package; *skewness* method) exhibited relatively constant abundances across subjects, fluctuating symmetrically around a mean value. The skewed types included both rare and prevalent bacteria; these exhibited left- and right-skewed types (skewness < -0.5 and >0.5, respectively). Four groups exhibited a left-skewed tail indicating low levels of such bacteria in some subjects. The 16 right-skewed types indicated notably high levels of these bacteria in some subjects. The 6 bimodal groups exhibited abundance profiles with two distinct states. To avoid potential biases from different DNA extraction methods, this categorization into abundance types was based on 401 unique subjects processed with the Repeated Bead Beating (RBB), a mechanical cell disruption method for DNA extraction that has been shown to be superior to alternatives in terms of DNA extraction efficiency[28].

**Bimodality detection with potential analysis** We screened the 130 genus-like groups targeted by the HITChip microarray for significance of bimodal abundance patterns across 1006 western adults using potential analysis[14]. This distribution-free approach is based on nonlinear dynamical systems theory, and was recently introduced to study analogous state shifts in climatology and forest ecosystems[13-14]. It is particularly suited for state identification in noisy dynamical systems. We assume an underlying stochastic system that has a potential function of the form

$$dz = - U'(z)dt + \sigma dW,$$

where $U(z)$ is the potential function, $z$ is the state variable (in our case the microbe log-abundance), $\sigma$ is the noise level and $dW$ is a noise term (Wiener process). The minima of the potential function correspond to stable states of the dynamical system. The corresponding Fokker-Planck equation connects the potential of this model to the probability density of the state variable $z$. The potential $U$ given by:

$$U = - \sigma \log(p_d)/2,$$

where $p_d$ is the empirically estimated probability density function of $z$. As our primary interest is in qualitative analysis of the potential function shape, we scaled the potential to the noise level $(U/\sigma^2)$[13]. We estimated the probability density with the R *density* function with Gaussian kernels and automated kernel width adjustment with the standard nrd method (R *stats* package)[13], giving the bandwidth $h=1.06 \ s/n^{1/5}$, where $s$ is the standard deviation of the data points and $n$ is the sample size. We adjusted this bandwidth with a factor of 0.8 to achieve more sensitive detection of multimodality. We estimated the local minima and maxima numerically and reduced spurious findings by requiring that the difference between local minima and the closest maxima was larger than a threshold value[13] (0.001 in our analysis) with a minimum density of 0.05. The number of local minima quantified the bi- or multimodality. We assessed state robustness by bootstrap analysis, where various subsets are selected from the original data by random sampling with replacement. This highlights the bacterial groups where bimodality is robust to variations induced by sampling. We repeated the analysis with 100 data sets obtained by bootstrapping, and selected the taxonomic groups that exhibited multimodal patterns in ≥80% of the cases. The evidence was the strongest for two (rather than three or more) alternative states in all multimodal groups. We determined the tipping point between the two states in these bimodal groups as the average over the corresponding bootstrap samples. We have implemented the potential analysis methodology of Livina *et al.*[14] in the *earlywarnings* R package (www.early-warning-signals.org; *livpotential_ews* function). The analysis of 130 genus-like groups revealed altogether six prevalent groups with bimodal abundance distributions: *Dialister spp.*, relatives of *Bacteroides fragilis, Prevotella melaninogenica* and *Prevotella oralis,* and bacteria belonging to the Uncultured Clostridiales groups I and II (UCI and UCII). Our sample collection included samples from multiple studies where different DNA extraction methods were used, including mechanic and enzymatic lysis protocols[28]. This can potentially affect the abundance estimates in a taxon-specific manner. To further control for this potential source of bias, we confirmed the bimodality in the subset of 401 samples with the RBB DNA extraction (see above). Without this extra verification, we observed bimodal patterns also in bifidobacteria and ruminococci that thus appear to be confounded by different extraction methods, highlighting the necessity to take DNA extraction method into account in meta-analyses that integrate data across independent studies[28] . Moreover, clostridia*,* Uncultured *Mollicutes,* and relatives of *Clostridium colinum* and *Clostridium difficile* showed indication of bimodality in the RBB subset, but we could not confirm the bimodality for these groups in the overall data collection. Finally, we verified the bimodality with the Prediction Strength index (PSI with M=20 random splits into training and test sets; *fpc* R package) based on the partition around medoids (PAM) clustering with Euclidean distance. While this combination was recently recommended to identify distinct community types[10], we applied it to assess bimodality in the individual bacterial groups. Also this approach supported our findings by providing strong evidence of bimodality (PSI>0.90) for all bimodal groups detected by potential analysis (*Dialister spp.*, UCI, UCII, and relatives of *P. oralis* and *P.melaninogenica)*, except *B.fragilis* which had moderate support (PSI>0.80) for bimodality.

We investigated in detail the six above-mentioned bimodal groups whose bimodality was consistently supported by all analyses. The high-abundance state was the least frequent for the *P.oralis* and *P.melaninogenica* groups (19% and 20% of the subjects, respectively) and *Dialister spp.* (28%)*,* while UCI (53%), UCII (57%), and the *B.fragilis* group (75%) were more often at the high-abundance state. The high-abundance *Prevotella* group had the most notable effect on the overall microbiota composition; using the total HITChip signal as a proxy for relative abundance, the high-abundance *P.melaninogenica* and *P.oralis* groups had an average relative abundance of 39.6% and 6.3%, respectively. This suggests that high-abundance *Prevotella* has a dominating role in intestinal microbiota. The other bimodal groups had more moderate effects on the overall composition, with a relative abundance of <2% on average. We also observed a substantial overlap between *B.fragilis* and *Prevotella*, with no significant differences in *B.fragilis* abundance between the low- and high-abundance *Prevotella* subjects (t-test; p>0.05 and fold-change <10%; see also Fig. 2). Similar result on the *Bacteroides* genus have been reported earlier[21] but are in contrast with other studies that have reported a notable tradeoff between *Bacteroides* and *Prevotella*[6,9,30].

**Community-level identification of alternative states** For comparison, we investigated evidence of alternative states at the overall ecosystem level. A variety of methodologies have been suggested to detect multimodality in microbiota profiling data sets, and it has been noted that community-level analysis of alternative states can be sensitive to the choice of dissimilarity measure, clustering method, and the approach for validating the number of clusters[10]. Popular choices for dissimilarity measures include Jensen-Shannon divergence, beta diversity measures, and standard correlation analyses. Partition Around Medoids (PAM) has been a frequent choice for a clustering method in microbiota profiling studies, and common methodologies for cluster number validation include the Kalinski-Harabasz index, Silhouette width, and Prediction strength[10]. We used the PAM clustering with Jensen-Shannon dissimilarity for community-level multivariate analysis of alternative states, coupled with the Prediction Strength approach to determine the optimal cluster number as this combination was earlier shown to have a good overall performance[10]. This revealed in our data two community-level alternative states, roughly corresponding to low and high *Prevotella* (relatives of *P.melaninogenica* and *P.oralis*).

**Cross-hybridization control** Cross-hybridization may cause bias in microarray analyses. We controlled this based on pre-calculated cross-alignment tables between the taxonomic groups targeted by HITChip. Cross-hybridization was negligible (<10%) between the bimodal groups and other taxa. *B.fragilis*, targeted by 40 probes, was an exception having 43% of shared probes with *B.ovatus et rel*. Since this could potentially contribute to the bimodality of the *B.fragilis* abundance distribution, we investigated the 16 probes that were specific for the *B.fragilis* group. Bimodal abundance patterns were detected in 25% of the unique probes, suggesting that this group may contain both bimodal and smoothly varying higher-level phylotypes. The highly correlated *P.oralis* and *P.melaninogenica* groups had 13-26% of shared probes. The correlation between these two groups remained high (r=0.81) after excluding the shared probes, however, confirming positive association.

**Associations with the host parameters** While the increased sample size obtained by combining data from multiple studies can improve statistical power, the potentially unbalanced distribution of host parameters across studies makes it potentially more difficult to distinguish between effects from the different sources. Hence, we quantified these associations by using subject metadata (age, body-mass index, sex, nationality) and DNA extraction method as fixed effects in multiple logistic regression to predict the state of the corresponding bacteria in each subject. The significance of each host parameter was estimated  by comparing the models with and without the corresponding parameter in predicting the state (the glm function in the R *stats* package). Missing annotations were omitted, and the false discovery rate was estimated to correct for multiple testing based on the Benjamini-Hochberg method using the *p.adjust* function (R *stats* package). We considered the associations with FDR<5% significant. Body-mass index was negatively associated with UCI and UCII; age positively associated with UCI and UCII, and negatively with *Dialister* and *Prevotella* (*melaninogenica and oralis* groups). Males had higher levels of *Prevotella*, and lower levels of *B.fragilis* compared to women. *Prevotella* was underrepresented in Nordic and Southern European countries and UCII in South Europe, while *Dialister* and *B.fragilis* were particularly common in Northern Europe (Scandinavia, UK and Ireland).

**Analysis of health associations** To investigate whether the alternative states are associated with health status, we compared the 1006 non-compromised adults to additional samples in the HITChip database from western adults with irritable bowel syndrome (106), metabolic syndrome (66), type 2 diabetes (78), ulcerative colitis (62), and cardiovascular disease (45). In addition, we divided the non-compromised subjects into severe obese (BMI≥35; N=136) and other (BMI<35; N=870) subjects. We used multiple logistic regression to predict disease occurrence from logarithmic abundance of each bimodal bacterial group and the potentially confounding variables DNA extraction method, age, body-mass index, and sex; 205 subjects (20.4%) for which DNA extraction method information was not available in the database were excluded from the analysis. The p-values were corrected with the Benjamini-Hochberg procedure. For exploratory purposes, associations with FDR<20% were considered to establish a trend, and FDR <5% to be significant (Supplementary Table 2). The UCI and UCII were negatively associated with severe obesity and Irritable Bowel Syndrome (IBS). Metabolic syndrome (MetS) was positively associated with the B. *fragilis* group and *Dialister*, and negatively with *Prevotella* (relatives of *P.oralis* and *P.melaninogenica)*. We compared the performance of the logistic regression model with and without bacterial abundance based on Receiver Operator Characteristic (ROC) analysis to detect disease occurrence; the average area-under-curve (AUC) values were derived from five-fold cross-validation. The differences in the AUC values between the models with and without bacterial abundance were small (0.1 increase with bacterial abundance) in the MetS group, indicating limited predictive ability in our data to distinguish the effects of MetS from the effects of age, sex and body-mass index. In severe obesity, inclusion of bacterial abundance notably improved disease detection (AUC 0.58 vs. 0.69 for UCI; 0.60 vs. 0.68 for UCII).

**Stability analysis** To investigate the temporal stability of the observed alternative states, we analyzed 78 subjects with additional follow-up data. The temporal range of the follow-up observations varied between one week and 36 months; most (95%) time points were within 1-9 months from the baseline. The overall temporal variability of bacterial abundance is illustrated for each bimodal group in Supplementary Fig. 2. We recorded the time from the baseline time point to the first observed state shift, and estimated the state shift frequencies as a function of time based on Kaplan-Meier survival analysis. The stability of each state, quantified by the fraction of subjects staying in the original state during the study interval, reflected the state frequencies in the overall data collection of 1006 subjects (Supplementary Fig. 5). All bimodal groups except *B.fragilis* appear more stable than expected based on empirical distributions derived from the data. The two *Prevotellas* mix less and are hence more stable than any other group, followed by UCI (less mixing than in 93% of the comparison groups), *Dialister* (83%) and UCII (78%). The relatively unstable *B.fragilis* group is an exception (35%). To control for potential effects of dietary intervention, affecting 28% of the samples in our follow-up data, we investigated differences between the intervention and other subjects in the bimodal groups. We did not observe significant differences between these two groups, except for UCI and UCII (t-test; p <0.05). This suggests that the analysis of the 78 subjects for short-term temporal stability of UCI and UCII may be affected by dietary intervention in our data, in line with the observation that the state shift frequencies in all bimodal groups except UCI and UCII approximately follow their stable state equilibrium (Supplementary Fig. 5).

**Species-level abundance distributions for the bimodal groups** Our analysis is based on genus-level classification. More refined differences between closely related bacteria may exist at the phylotype level. The species-level phylotypes exhibited differences in their bimodality patterns. The most strikingly bimodal bacteria were uncultured phylotypes, as in the case of *P.melaninogenica* and *P.oralis* groups, where the type species were detected only at low abundance, but the related uncultured bacteria exhibited strongly bimodal distributions. We observed notable variation within the genus *Prevotella* as *P. ruminicola* exhibited only weak correlation with the combined *P.oralis / P.melaninogenica axis* (r=0.39). The bimodally distributed cultured species included *B. nordii* (*B.fragilis* group) and *D. invisus* (*Dialister spp.*). The heterogeneous abundance profiles of individual phylotypes within each bimodal genus-like group could indicate potential antagonism or metabolic specialization between the implicated phylotypes. For the uncultured UCI and UCII groups that do not include any cultured representatives we performed a 16S rRNA target gene sequences alignment against the RDP database[29] to update their phylogetic assignments. The majority (19/30) of the phylotypes within the UCI group were identified as members of family Ruminococcaceae, and 12 could be further assigned to genus Acetivibrio (Supplementary Table 3). Uncultured bacterium clones Eldhufec 308 and Eldhufec312, showed clear bimodal abundance distribution. Similar abundance distribution pattern characterized also the phylotypes within the UCII group, which could not be identified further down than order Clostridiales.

**Acknowledgments:** We thank Outi Immonen, Philippe Puylaert, and Jarmo Ritari for technical support with the HITChip data analysis, and Vasilis Dakos and Erwin Zoetendal for their feedback. This work was supported by Academy of Finland [256950 to LL; 141140 to WMdV]; European Research Council [ERC250172 to WMdV; ERC268732 to MS]; Spinoza Award, Netherlands Organization for Scientific Research [2008 to WMdV;  2009 to MS]; Alfred Kordelin foundation [to LL]. Correspondence and requests for materials should be addressed to leo.lahti@ïki.fi.

**Author contributions** LL, MS and WMdV designed the study. LL carried out data processing and statistical analysis, coordinated the study and prepared the manuscript. AS, LL, JS, and WMdV contributed to acquisition of data. JS contributed to data analysis. All authors contributed to interpreting the results and writing the manuscript.

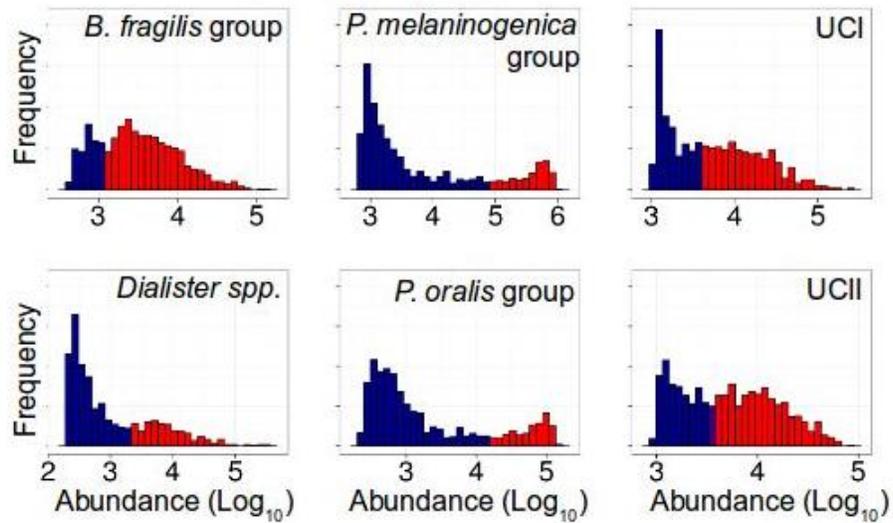

**Fig. 1 The bimodal bacteria** Logarithmic abundance distributions of the six bimodal phylogenetic groups that exhibit robust alternative states of low (blue) and high (red) abundance across intestinal microbiota of a thousand western adults[15]. The UCI and UCII refer to the Uncultured Clostridiales I and II, respectively. The frequency of the observations is shown as a function of the phylogenetic microarray signal[12].

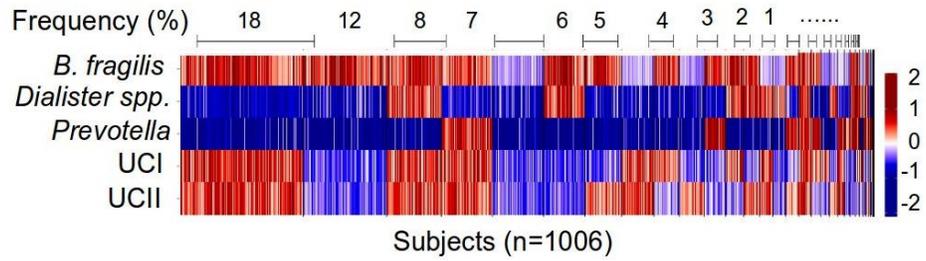

**Fig. 2 The tipping elements co-occur in various combinations** The bimodal groups (rows) co-occur in various combinations within the study population. Shading indicates the relative abundance ($\log_{10}$) for each group with respect to the identified tipping point between the alternative states of low (blue) and high (red) abundance. The subjects are ordered by the distinct combinations. The most frequent combination (18%) is the high-abundance UCI, UCII, and *B.fragilis* groups with low-abundance *Dialister* and *Prevotella*.

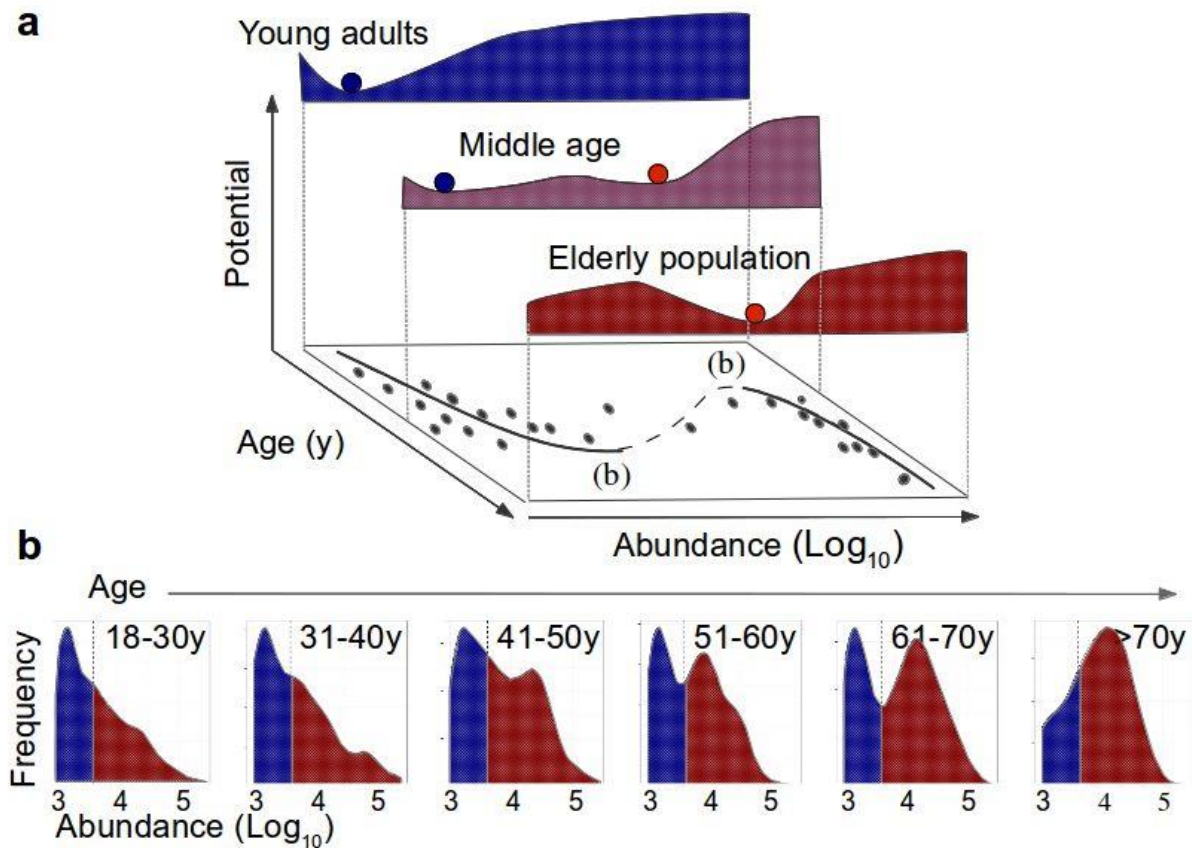

**Fig. 3 State shifts associated with ageing A** A schematic illustration of UCI state shift during ageing. The UCI abundance data (bottom plane) suggests a catastrophe-fold, where the solid and dashed parts of the curve correspond to stable states and unstable equilibria, respectively. The depth and width of the potential minima indicate decreasing resilience of the system towards the bifurcation points during ageing (b). **B** Observation density in various age groups highlights the associations between UCI abundance and age. The alternative states are more clearly pronounced than in the population-level histogram pooled over all age groups (Fig. 1).

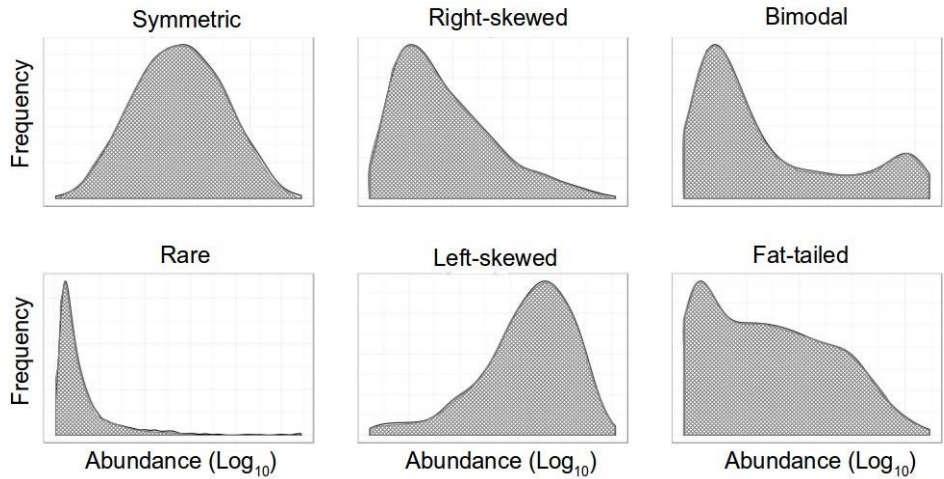

**Supplementary Fig. 1 Bacterial abundance types** Bacterial abundance types include symmetric, skewed and bimodal abundance distributions (Supplementary Table 1). The skewed types include prevalent left- and right-skewed types as well as rare bacteria. The bimodal types include cases with two distinct peaks, or with a peak at low abundance combined with a more widely varying fat tail of high-abundance subjects (Fig. 1).

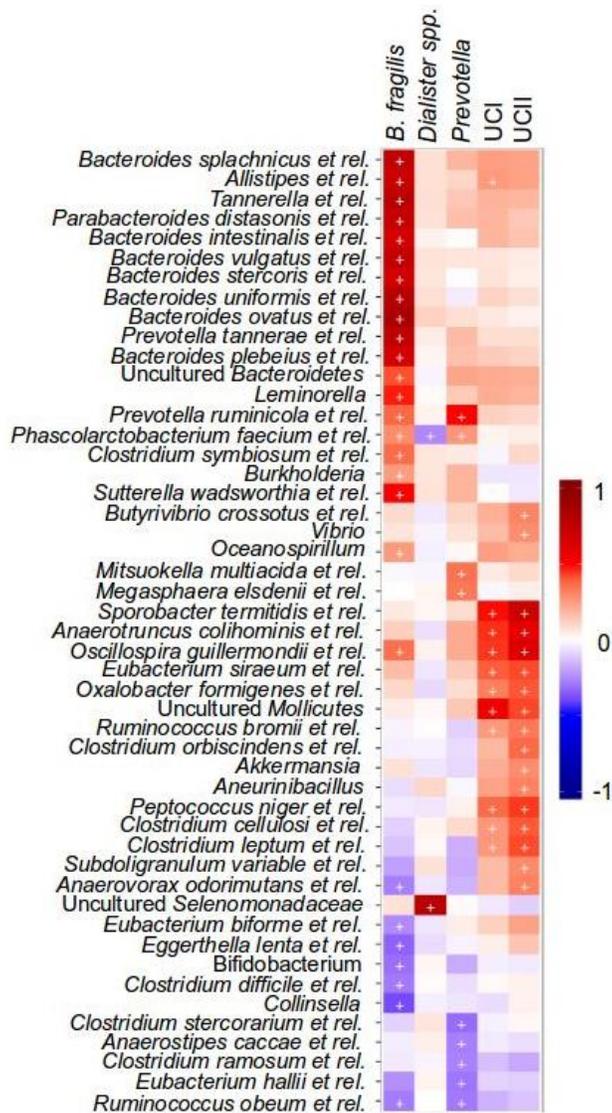

**Supplementary Fig. 2 Temporal variation of the bimodal groups** Temporal variation in each bimodal group during the follow-up interval (1-9 months from the baseline for 95% of the subjects[15]). Each horizontal line indicates the abundance range of the bimodal group for one subject during the follow-up period. The 78 subjects are ordered based on their mean abundance. The red lines highlight subjects visiting the alternative state during the study period (Supplementary Fig. 6). The tipping point between the two states, estimated based on 1006 subjects, is indicated for each group by the solid vertical line.

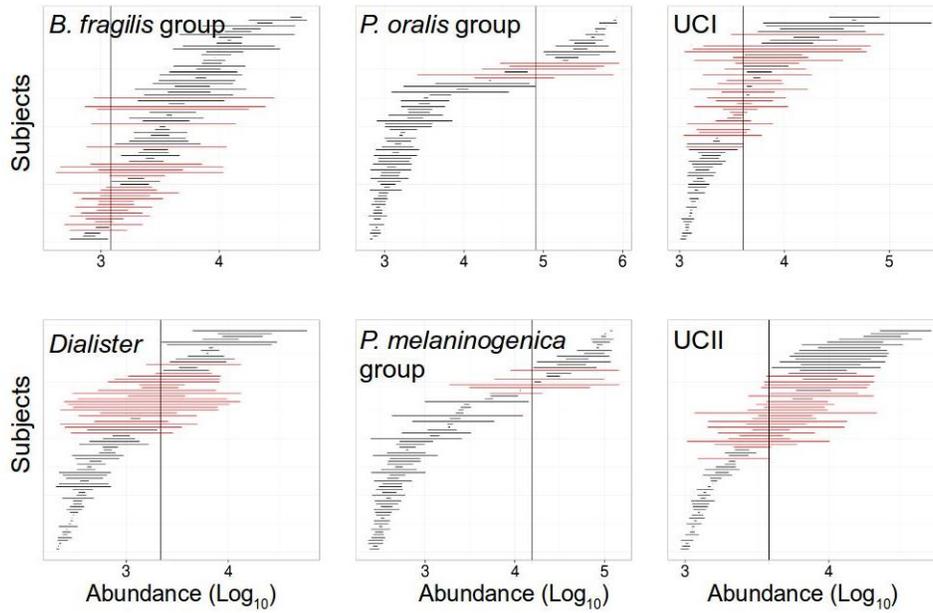

**Supplementary Fig. 3 Correlation between the bimodal groups and other bacteria**
Correlation between the bimodal groups (columns) and the 124 other genus-like groups targeted by the HITChip microarray (rows); 75 groups did not show remarkable correlations (|r|>0.25) with the bimodal groups and have been removed for clarity. The shading indicates the Pearson correlation coefficient; significant correlations (FDR<5%) with 49 genus-like groups are indicated by '+'.

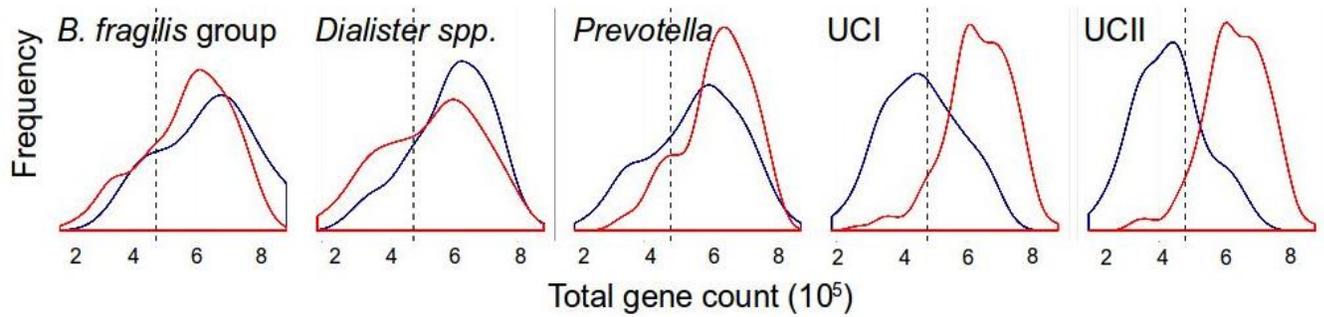

**Supplementary Fig. 4 Alternative states and gene richness** Total gene count for 255 subjects (39-72 years) in a metagenomic sequencing study[17], where a bimodal gene count was reported (the modes divided by the vertical dashed line). The total gene counts are illustrated for the subjects with low (blue) and high (red) abundance of each bimodal group. The total gene count is significantly associated with all groups ($p < 0.01$; Wilcoxon test) except *B.fragilis* ($p=0.2$). *B.fragilis* and *Dialister spp.* were associated with the low gene count[17]; *Prevotella* (relatives of *P.melaninogenica* and *P.oralis)*, UCI and UCII were associated with high gene count.

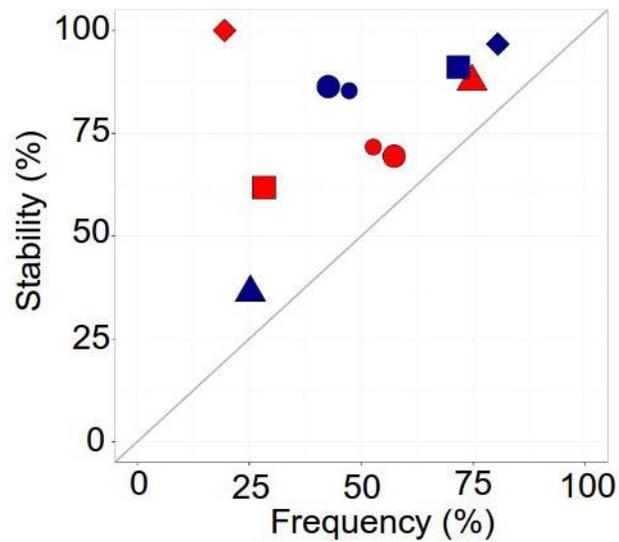

**Supplementary Fig. 5 Temporal stability of the alternative states** State frequency in the study population (horizontal axis; n=1006) predicts short-term stability for the 78 follow-up subjects over a three-month interval (vertical axis). The high-abundance *Prevotella* (♦) is an exception as it is both the least frequent and the most stable state. All groups appear more stable than expected (gray line). Symbols: *B.fragilis* group (▲), *Dialister spp.* (■), *Prevotella* (♦), UCI (•), UCII (●); the low- and high-abundance states are indicated by blue and red, respectively.

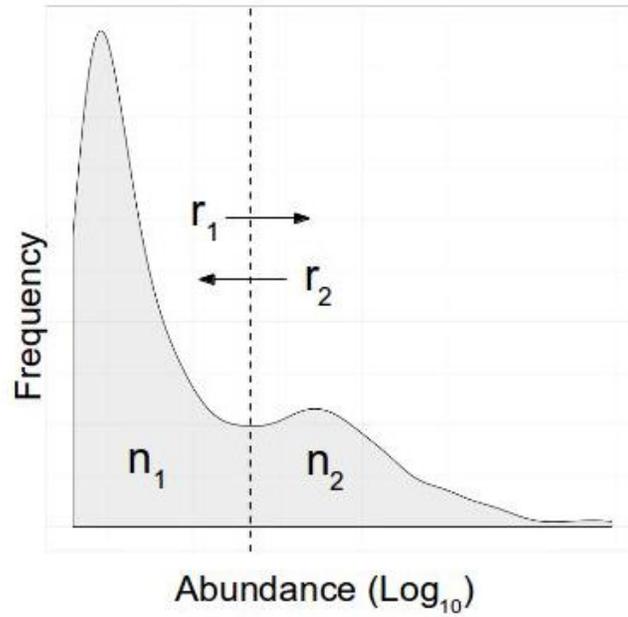

**Supplementary Fig. 6 State shifts between the alternative states** State frequency predicts state stability. In stationary state the ratio of the switching rates $r_i$ should be inversely related to state frequencies n ($r_1/r_2 = n_2/n_1$) assuming a continuous-time Markov process.

| | |
|---|---|
| Symmetric (34) | *Akkermansia; Allistipes et rel.; Anaerostipes caccae et rel.; Anaerovorax odorimutans et rel.; Bacteroides ovatus et rel.; Bacteroides plebeius et rel.; Bacteroides uniformis et rel.; Bacteroides vulgatus et rel.; Bifidobacterium; Bryantella formatexigens et rel.; Clostridium cellulosi et rel.; Clostridium leptum et rel.; Clostridium nexile et rel.; Clostridium orbiscindens et rel.; Clostridium sphenoides et rel.; Clostridium stercorarium et rel.; Clostridium symbiosum et rel.; Coprococcus eutactus et rel.; Dorea formicigenerans et rel.; Eubacterium hallii et rel.; Eubacterium rectale et rel.; Eubacterium ventriosum et rel.; Lachnobacillus bovis et rel.; Oscillospira guillermondii et rel.; Outgrouping clostridium cluster XIVa; Oxalobacter formigenes et rel.; Papillibacter cinnamivorans et rel.; Roseburia intestinalis et rel.; Ruminococcus bromii et rel.; Ruminococcus callidus et rel.; Ruminococcus gnavus et rel.; Ruminococcus obeum et rel.; Sporobacter termitidis et rel.; Tannerella et rel.* |
| Left-skewed (4) | *Butyrivibrio crossotus et rel.; Faecalibacterium prausnitzii et rel.; Lachnospira pectinoschiza et rel.; Subdoligranulum variable et rel.* |
| Right-skewed (16) | *Anaerotruncus colihominis et rel.; Bacteroides fragilis et rel.; Bacteroides splachnicus et rel.; Bacteroides stercoris et rel.; Clostridia; Clostridium difficile et rel.; Collinsella; Eubacterium biforme et rel.; Lactobacillus plantarum et rel.; Parabacteroides distasonis et rel.; Prevotella tannerae et rel.; Ruminococcus lactaris et rel.; Streptococcus bovis et rel.; Streptococcus mitis et rel.; Sutterella wadsworthia et rel.; Uncultured Mollicutes* |
| Bimodal (6) | *Dialister; Prevotella melaninogenica et rel.; Prevotella oralis et rel.; Uncultured Clostridiales I; Uncultured Clostridiales II, Bacteroides fragilis et rel.* |
| Rare (70) | *Actinomycetaceae; Aerococcus; Aeromonas; Alcaligenes faecalis et rel.; Anaerobiospirillum; Anaerofustis; Aneurinibacillus; Aquabacterium; Asteroleplasma et rel.; Atopobium; Bacillus; Bacteroides intestinalis et rel.; Bilophila et rel.; Brachyspira; Bulleidia moorei et rel.; Burkholderia; Campylobacter; Catenibacterium mitsuokai et rel.; Clostridium felsineum et rel.; Clostridium ramosum et rel.; Clostridium thermocellum et rel.; Coprobacillus catenaformis et rel.; Corynebacterium; Desulfovibrio et rel.; Eggerthella lenta et rel.; Enterobacter aerogenes et rel.; Enterococcus; Escherichia coli et rel.; Eubacterium cylindroides et rel.; Eubacterium limosum et rel.; Eubacterium siraeum et rel.; Fusobacteria; Gemella; Granulicatella; Haemophilus; Helicobacter; Klebisiella pneumoniae et rel.; Lactobacillus catenaformis et rel.; Lactobacillus gasseri et rel.; Lactobacillus salivarius et rel.; Lactococcus; Leminorella; Megamonas hypermegale et rel.; Megasphaera elsdenii et rel.; Methylobacterium; Micrococcaceae; Mitsuokella multiacida et rel.; Moraxellaceae; Novosphingobium; Oceanospirillum; Peptococcus niger et rel.; Peptostreptococcus anaerobius et rel.; Peptostreptococcus micros et rel.; Phascolarctobacterium faecium et rel.; Prevotella ruminicola et rel.; Propionibacterium; Proteus et rel.; Pseudomonas; Serratia; Staphylococcus; Streptococcus intermedius et rel.; Uncultured Bacteroidetes; Uncultured Chroococcales; Uncultured Selenomonadaceae; Veillonella; Vibrio; Weissella et rel.; Wissella et rel.; Xanthomonadaceae; Yersinia et rel.* |

**Supplementary Table 1. Bacterial abundance types** Abundance types of the 130 genus-like bacterial groups quantified by the phylogenetic HITChip microarray. The genus-like phylogenetic groups (>90% sequence similarity in the 16S rRNA gene) are referred to as *type species* and relatives, the latter being shortened as "*et rel.*"[12] The genus-like groups are categorized into symmetric, skewed, bimodal, and rare groups[15].

| Phylum | Genus-like group | Species-level phylotypes | Accession number |
|---|---|---|---|
| Bacteroidetes | Bacteroides fragilis et rel. | bacterium adhufec20 | AF132255 |
| | | bacterium adhufec305 | AF132245 |
| | | Bacteroides thetaiotaomicron | L16489 |
| | | Bacteroides fragilis | M11656 |
| | | Uncultured bacterium NR59 | AY916213 |
| | | Uncultured bacterium ZN01 | AY916174 |
| | | Uncultured bacterium LC37 | AF408304 |
| | | Uncultured bacterium LCFC73 | AF364892 |
| | | Bacteroides fragilis | X83935 |
| | | Bacteroides ovatis | X83957 |
| | | Bacteroides caccae | X83956 |
| | Prevotella melaninogenica et rel. | bacterium adhufec203 | AF132246 |
| | | bacterium adhufec45 | AY920161 |
| | | Prevotella albensis | AJ011683 |
| | | Prevotella intermedia | AF414621 |
| | | Prevotella salivae | AB108826 |
| | | Prevotella sp. Br 64 | AJ581354 |
| | | Prevotella sp. CB7 | AB547423 |
| | | Prevotella melaninogenica | L16469 |
| | | Uncultured bacterium clone Eldhufec038 | AY916983 |
| | | Uncultured bacterium clone Eldhufec037 | AY916982 |
| | | Uncultured bacterium clone Eldhufec038 | AY916913 |
| | | Uncultured bacterium clone Eldhufec038 | AY916912 |
| | | Uncultured bacterium clone Eldhufec038 | AY916911 |
| | | Uncultured bacterium clone Eldhufec038 | AY916910 |
| | | Uncultured bacterium B178 | AY916316 |
| | | Uncultured bacterium NI17 | AY916144 |
| | | Uncultured bacterium OLDC-A6 | AB089782 |
| | | Uncultured Prevotella sp. NB039 | AB064832 |
| | Prevotella oralis et rel. | Prevotella buccalis | L16476 |
| | | Prevotella oralis | L16480 |
| | | Prevotella sp. CB25 | AB547424 |
| | | Uncultured bacterium HuCC28 | AJ515453 |
| | | Uncultured bacterium clone Eldhufec011 | AY916986 |
| | | Uncultured bacterium clone Eldhufec011 | AY916915 |
| Firmicutes, Clostridium cluster XI | Dialister | Dialister invisus | AY162469 |
| | | Dialister pneumosintes | X82500 |
| | | Uncultured Gram-positive bacterium NIUB11 | AB061409 |
| | | Uncultured bacterium clone Eldhufec038 | AY916969 |
| | | Uncultured bacterium clone Eldhufec038 | AY916964 |
| | | Uncultured bacterium M510 | AY892155 |
| Firmicutes, Uncultured Clostridiales | Uncultured Clostridiales I | Uncultured human gut bacterium JW084 | AB500932 |
| | | Uncultured human gut bacterium JW0R12 | AB500931 |
| | | Uncultured bacterium CLD4-C7 | AB069786 |
| | | Uncultured bacterium CLD4-F6 | AB069785 |
| | | Uncultured bacterium CLD4-F7 | AB069784 |
| | | Uncultured bacterium OLD4-H6 | AB069792 |
| | | Uncultured bacterium OLD6-N6 | AB069790 |
| | | Uncultured bacterium OLDC4-1 | AB069791 |
| | | Uncultured bacterium C116 | AY916329 |
| | | Uncultured bacterium C257 | AY916329 |
| | | Uncultured bacterium C427 | AY916343 |
| | | Uncultured bacterium CK46 | AY916352 |
| | | Uncultured bacterium D279 | AY916363 |
| | | Uncultured bacterium D650 | AY916381 |
| | | Uncultured bacterium LI465 | AY916259 |
| | | Uncultured bacterium M203 | AY916153 |
| | | Uncultured bacterium M223 | AY916151 |
| | | Uncultured bacterium M412 | AY916155 |
| | | Uncultured bacterium M621 | AY916183 |
| | | Uncultured bacterium MF20 | AY916236 |
| | | Uncultured bacterium MF36 | AY916234 |
| | | Uncultured bacterium MG89 | AY916261 |
| | | Uncultured bacterium NI499 | AY916173 |
| | | Uncultured bacterium clone Eldhufec038 | AY920163 |
| | | Uncultured bacterium clone Eldhufec313 | AY920167 |
| | | Uncultured bacterium clone Eldhufec314 | AY920169 |
| | | Uncultured bacterium clone Eldhufec317 | AY920151 |
| | | Uncultured bacterium clone Eldhufec317 | AY920152 |
| | | Uncultured bacterium UC7-9 | AJ408307 |
| | | Uncultured bacterium UC7-107 | AJ408349 |
| | Uncultured Clostridiales II | Uncultured human gut bacterium JW0H13 | AB500990 |
| | | Uncultured bacterium OLD9-C3 | AB069776 |
| | | Uncultured bacterium C736 | AY916344 |
| | | Uncultured bacterium LQ85 | AY916263 |
| | | Uncultured bacterium M501 | AY916160 |
| | | Uncultured bacterium clone Eldhufec320 | AY920208 |
| | | Uncultured bacterium clone Eldhufec322 | AY920347 |
| | | Uncultured bacterium cathufac16c6 | AF520251 |
| | | Uncultured bacterium cathufac17d6 | AF520240 |
| | | Uncultured bacterium OLD8-H1 | AB069779 |
| | | Uncultured bacterium OLD8-F4 | AB069777 |
| | | Uncultured bacterium OLDC-A2 | AB069780 |
| | | Uncultured bacterium C360 | AY916358 |
| | | Uncultured bacterium C606 | AY916361 |
| | | Uncultured bacterium D191 | AY916361 |
| | | Uncultured bacterium K542 | AY916185 |
| | | Uncultured bacterium M403 | AY916155 |
| | | Uncultured bacterium MH87 | AY916256 |
| | | Uncultured bacterium MM64 | AY916264 |
| | | Uncultured bacterium HuC96 | AJ408362 |
| | | Uncultured bacterium clone Eldhufec326 | AY920203 |
| | | Uncultured bacterium clone Eldhufec326 | AY920204 |
| | | Uncultured bacterium clone Eldhufec326 | AY920205 |
| | | Uncultured bacterium clone Eldhufec326 | AY920211 |
| | | Uncultured human gut bacterium JW1H11 | AB500931 |
| | | Uncultured human gut bacterium JW1B0 | AB500979 |

**Supplementary Table 2 Phylotype-level characterization of the bimodal groups** Cultivated species and uncultured phylotypes (≥98% 16S rRNA gene sequence similarity) that constitute the six bimodal genus-like phylogenetic groups on the HITChip microarray[12].

**Supplementary Table 3. Phylogenetic characterization of the UCI and UCII groups**

| UCI | RDP sequence | HITChip phylotype name | Accession |
|---|---|---|---|
| | Firmicutes; Clostridia; Clostridiales; Ruminococcaceae; Acetivibrio | | |
| | uncultured bacterium; C118 | uncultured bacterium C118 | AY916326 |
| | uncultured bacterium; Eldhufec308 | Uncultured bacterium clone Eldhufec308 | AY920183 |
| | uncultured bacterium; clone Eldhufec312 | uncultured bacterium clone Eldhufec312 | AY920187 |
| | uncultured bacterium; clone Eldhufec314 | Uncultured bacterium clone Eldhufec314 | AY920189 |
| | uncultured bacterium; Eldhufec317 | Uncultured bacterium clone Eldhufec317 | AY920192 |
| | uncultured bacterium; D049 | uncultured bacterium D049 | AY916352 |
| | uncultured bacterium; M220 | uncultured bacterium M220 | AY916150 |
| | uncultured bacterium; MF22 | uncultured bacterium MF22 | AY916236 |
| | uncultured bacterium; MF35 | uncultured bacterium MF35 | AY916239 |
| | uncultured bacterium; NH06 | uncultured bacterium NH06 | AY916173 |
| | Firmicutes; Clostridia; Clostridiales; Ruminococcaceae; unclassified Ruminococcaceae | | |
| | uncultured bacterium; Eldhufec316 | Uncultured bacterium clone Eldhufec316 | AY920191 |
| | uncultured bacterium; C257 | uncultured bacterium C257 | AY916329 |
| | uncultured bacterium; C627 | uncultured bacterium C627 | AY916340 |
| | uncultured bacterium; M233 | uncultured bacterium M233 | AY916151 |
| | uncultured bacterium; M412 | uncultured bacterium M412 | AY916156 |
| | uncultured bacterium; MG86 | uncultured bacterium MG86 | AY916291 |
| | uncultured bacterium; OLDA-F6 | uncultured bacterium OLDA-F6 | AB099785 |
| | uncultured bacterium; UC7-127 | Uncultured bacterium UC7-127 | AJ606249 |
| | uncultured bacterium; UC7-9 | Uncultured bacterium UC7-9 | AJ606227 |
| | Firmicutes; Clostridia; Clostridiales; unclassified Clostridiales | | |
| | uncultured bacterium; D279 | uncultured bacterium D279 | AY916363 |
| | uncultured bacterium; D693 | uncultured bacterium D693 | AY916361 |
| | uncultured bacterium; LH65 | uncultured bacterium LH65 | AY916206 |
| | uncultured bacterium; DGGE gel band; 128-BL-00-B1 | uncultured bacterium M621 | AY743526 |
| | uncultured bacterium; OLDA-C7 | uncultured bacterium OLDA-C7 | AB099786 |
| | uncultured human intestinal bacterium; JW2B4 | uncultured human gut bacterium JW2B4 | AB080852 |
| | uncultured human intestinal bacterium; JW2F12 | uncultured human gut bacterium JW2F12 | AB080851 |
| | Firmicutes; unclassified Firmicutes | | |
| | uncultured bacterium; OLDB-A9 | uncultured bacterium OLDB-A9 | AB099783 |
| | uncultured bacterium; OLDA-H9 | uncultured bacterium OLDA-H9 | AB099782 |
| | unclassified Bacteria | | |
| | uncultured bacterium; OLDC-A1 | uncultured bacterium OLDCA-1 | AB099781 |
| | uncultured bacterium; OLDA-F7 | uncultured bacterium OLDA-F7 | AB099784 |
| **UCII** | **RDP sequence** | **HITChip phylotype name** | **Accession** |
| | Firmicutes; Clostridia; Clostridiales; unclassified Clostridiales | | |
| | uncultured bacterium; C583 | uncultured bacterium C583 | AY916338 |
| | uncultured bacterium; D191 | uncultured bacterium D191 | AY916361 |
| | uncultured bacterium; K342 | uncultured bacterium K342 | AY916195 |
| | uncultured bacterium; LQ86 | uncultured bacterium LQ86 | AY916269 |
| | uncultured bacterium; MH67 | uncultured bacterium MH67 | AY916298 |
| | uncultured bacterium; MM92 | uncultured bacterium MM92 | AY916304 |
| | uncultured bacterium; MS01 | uncultured bacterium MS01 | AY916160 |
| | uncultured human intestinal bacterium; JW2H12 | uncultured human gut bacterium JW2H12 | AB080880 |
| | uncultured human intestinal bacterium; JW1H11 | uncultured human gut bacterium JW1H11 | AB080881 |
| | uncultured bacterium; OLDB-F4 | uncultured bacterium OLDB-F4 | AB099777 |
| | uncultured bacterium; OLDB-H1 | uncultured bacterium OLDB-H1 | AB099779 |
| | uncultured bacterium; M403 | uncultured bacterium M403 | AY916155 |
| | uncultured bacterium; Eldhufec328 | Uncultured bacterium clone Eldhufec328 | AY920203 |
| | uncultured bacterium; Eldhufec329 | Uncultured bacterium clone Eldhufec329 | AY920204 |
| | uncultured bacterium; sd56 | Uncultured bacterium clone Eldhufec333 | EU201635 |
| | uncultured bacterium; Eldhufec334 | Uncultured bacterium clone Eldhufec334 | AY920209 |
| | uncultured bacterium; Eldhufec336 | Uncultured bacterium clone Eldhufec336 | AY920211 |
| | uncultured bacterium; C655 | uncultured bacterium C655 | AY916341 |
| | uncultured bacterium; cadhufec1705sav | uncultured bacterium cadhufec1705 | AF530343 |
| | uncultured bacterium; cadhufec18c08sav | uncultured bacterium cadhufec18c08 | AF530351 |
| | Firmicutes; unclassified Firmicutes | | |
| | uncultured bacterium; OLDC-A2 | uncultured bacterium OLDC-A2 | AB099780 |
| | uncultured bacterium; Eldhufec332 | Uncultured bacterium clone Eldhufec332 | AY920207 |
| | unclassified Bacteria | | |
| | uncultured bacterium; 128g07 | uncultured bacterium C736 | AJ812148 |
| | uncultured bacterium; OLDB-C2 | uncultured bacterium OLDB-C2 | AB099778 |
| | uncultured human intestinal bacterium; JW1B2 | uncultured human gut bacterium JW1B2 | AB080879 |
| | uncultured bacterium; from human colonic sample; HuCA6 | uncultured bacterium HuCA6 | AJ408962 |

**Supplementary Table 3. Phylogenetic characterization of the UCI and UCII groups**

Ribosomal Database Project (RDP) alignment for the 16S rRNA target sequences of Uncultured Clostridiales I and II (UCI and UCII) detected with the HITChip microarray (October 2013). The RDP match score is 1 for all HITChip phylotypes except the UCII Uncultured bacterium clone Eldhufec333 (0.963). All sequences belong to the domain Bacteria. The RDP sequences are grouped according to their Phylum, Class, Order, Family and Genus; these are indicated up to the highest accessible taxonomic level, separated by semicolon

| Health status | Bimodal group | Enriched state | Compromised (%) | Controls (%) | FDR (%) |
|---|---|---|---|---|---|
| Severe obesity (n=136) | UCI | Low abundance | 29 | 55 | <0.1 |
| Severe obesity | UCII | Low abundance | 38 | 61 | <0.1 |
| IBS (n=106) | UCII | Low abundance | 50 | 61 | 1 |
| MetS (n=66) | *B.fragilis* group | High abundance | 89 | 78 | <0.1 |
| MetS | *Prevotella* group | Low abundance | 11 | 22 | 11 |
| MetS | *Dialister* | High abundance | 36 | 28 | 13 |

**Supplementary Table 4. Associations between health status and the bimodal bacterial groups** The third column ('Enriched state') indicates the state associated with the disease. The columns 4 and 5 show the proportion of subjects in the high-abundance state in the compromised and the healthy controls, respectively. To control for differences in subject characteristics or sample treatment between the two groups, we estimated the significance based on multiple logistic regression corrected for age, sex, body-mass index, and DNA extraction method, followed by Benjamini-Hochberg correction. The associations with FDR<20% are shown.